\documentclass[a4paper, 11pt]{article}

\usepackage[margin=25mm]{geometry}
\usepackage{amsmath}
\usepackage{graphicx} 
\usepackage{parskip} 
\usepackage{hyperref}
\usepackage{titlesec}
\usepackage[labelfont=bf,font=small]{caption}
\usepackage{subcaption}
\usepackage{float}
\usepackage[numbers,sort&compress]{natbib}
\usepackage{booktabs}

\hypersetup{
linktocpage,                                      
colorlinks  = true,                               
urlcolor    = cyan,
linkcolor   = red,
citecolor   = blue
}

\titleformat{\section}{\normalsize \bfseries}{\thesection}{1em}{}
\titleformat{\subsection}{\small \bfseries}{\thesubsection}{1em}{}
\titleformat{\subsubsection}{\small \bfseries}{\thesubsubsection}{1em}{}

\providecommand{\keywords}[1]
{
  \small
  \noindent \textbf{Keywords:} #1
}

\title{\vspace{-2cm} \rule{\textwidth}{1.5pt} \textbf{\LARGE Graph-Based Generalization of Galam Model: Convergence Time and Influential Nodes\vspace{-3mm}} \rule{\textwidth}{1.5pt}}

\author{\small \textbf{Sining Li} \vspace{-1mm} 
\\ \small School of Computing \vspace{-1mm} 
\\ \small Australian National University \vspace{-1mm}
\\ \small Canberra, ACT 2601, Australia \vspace{-1mm}
\\ \small sining.li@anu.edu.au \and 
\hspace{1.5cm} \small \textbf{Ahad N. Zehmakan} \vspace{-1mm}
\\ \hspace{1.5cm} \small School of Computing \vspace{-1mm}
\\ \hspace{1.5cm} \small Australian National University \vspace{-1mm}
\\ \hspace{1.5cm} \small Canberra, ACT 2601, Australia \vspace{-1mm}
\\ \hspace{1.5cm} \small ahadn.zehmakan@anu.edu.au }

\date{}

\begin{document}

\maketitle

\vspace{3mm}
\begin{abstract}
    \noindent We study a graph-based generalization of the Galam opinion formation model. Consider a simple connected graph which represents a social network. Each node in the graph is colored either blue or white, which indicates a positive or negative opinion on a new product or a topic. In each discrete-time round, all nodes are assigned randomly to groups of different sizes, where the node(s) in each group form a clique in the underlying graph. All the nodes simultaneously update their color to the majority color in their group. If there is a tie, each node in the group chooses one of the two colors uniformly at random. 
    
    \noindent Investigating the convergence time of the model, our experiments show that the convergence time is a logarithm function of the number of nodes for a complete graph and a quadratic function for a cycle graph. We also study the various strategies for selecting a set of seed nodes to maximize the final cascade of one of the two colors, motivated by viral marketing. We consider the algorithms where the seed nodes are selected based on the graph structure (nodes' centrality measures such as degree, betweenness, and closeness) and the individual's characteristics (activeness and stubbornness). We provide a comparison of such strategies by conducting experiments on different real-world and synthetic networks.
\end{abstract}



\section{Introduction}
\label{intro}

Humans' decisions and opinions can be influenced by others. When forming an opinion or making a decision on a subject such as a new product or a political election, people usually seek advice from their families, friends, and other people with whom they often keep in contact with. In addition, people also consider the opinions of the figures that they value or respect; for instance, figures in specific fields or celebrities that people look up to. Therefore, people often keep updating their decisions and opinions by interacting with their connections.

Furthermore, the prevalence of online social media and social networking platforms, such as Facebook, Instagram, and WeChat, contribute to the increasing pace of opinion formation. These online platforms provide people with a convenient and fast way to communicate with friends, obtain information, and express opinions. Consequently, they significantly accelerate the information spreading and opinion formation and exchange process.

In recent decades, there has been a growing demand for a deeper understanding of how opinions form and how information spreads in social networks. A variety of opinion diffusion models, which aim to simulate the process of information spreading in a society, have thus been developed and studied within sociophysics using tools and concepts from statistical physics~\cite{Yin2019}. More recently, the topic has been enriched by contributions from theoretical computer science~\cite{Bredereck2017, Huang2013}. 

The studied models typically involve a graph $G$ and an initial coloring of its nodes, with each node being either blue or white. The graph represents a social network, in which each node corresponds to an individual. The edges of the graph represent relationships between individuals, such as personal contacts, friendships, or followers. The color assigned to each node, say white or blue, represents the individual's positive or negative opinion on a subject. Following the initialization, a group of nodes in the graph update their colors based on a predefined rule in each round.

The majority model is an instance of the aforementioned basic model~\cite{Auletta2015, Auletta2018} and it has become popular within the community of social networks. In this model, all nodes in the graph $G$ simultaneously update their color to the most frequent color among their neighbors. If there is a tie, a node does not change its current color. It should be noted that the number of neighbors of each node can be quite large, which makes the model somewhat unrealistic. For example, it is almost impossible for a person who has 50 friends to discuss a topic with all the friends at the same time.

Another established model is the Galam model~\cite{Galam2002}, which describes the dynamics of the spreading of the minority opinion in a social debate. In this model, all individuals (nodes) are randomly assigned to groups of different sizes, based on the geometry of social life of people, with the discussing groups composed of small sizes from 1 person up to 5 or 6 people. Only the people in the same group can discuss the topic and update their opinion based on the local majority opinion in the group. Therefore, the Galam model is more realistic. In the general majority-based model, the nodes with a higher degree can influence more nodes in each round. But, in the Galam model, the node can only influence other nodes in the same group. Moreover, at a tie, the Galam model introduces the possibility of a tie breaking determined using the unconscious prejudices of the agents~\cite{Galam2005}.

In the classic Galam model, each individual has the same probability of grouping with any other individual. However, it is unusual that two people get together and discuss on some topics if they do not have any relationships. To add another layer of realism to the model, we leverage the underlying graph structure of a social network to introduce a generalization of the Galam model in the present paper. In the Graph-based Galam model (GGM), only the nodes that form a clique in the graph can be assigned to a group.

In the present paper, we focus on two problems: (1) How long does it take for the GGM to converge? This is arguably the most well-studied question 
for different dynamic systems. In our model, the answer highly depends on the underlying graph structure. (2) What is the subset of nodes which maximizes the final expected number of blue nodes, if initially blue, for a given fixed budget? This is also another popular question investigated for various models, motivated from viral marketing. For example, when a company plans to launch a new product on to the market, it is often intended to persuade a specific group of people into using the product to maximize the final cascade of the adoptions of the product. Another example is when a political party aims to convince a subset of individuals to adopt a positive opinion about a topic with the aim of a large further adoption of the positive opinion. The main problem here is to find a subset of nodes (individuals) of a given size whose adoption of the desired product/opinion results in its maximum adoption at the end.

We study the convergence time of GGM on different synthetic and real-world graph data. Experiments conducted on cycle and complete graphs show that the convergence times of GGM on cycle and complete graphs are in $\mathcal{O}(n^2)$ and $\mathcal{O}(\log n)$, respectively, where $n$ denotes the number of nodes. We also investigate how adding new edges can influence the convergence time of the process and observe that it essentially depends on the manner that the edges are added. Particularly, if the added edges increase the expansion (connectivity) of the graph, the convergence time usually decreases.

Furthermore, the experiments on complete graphs and real-world social networks indicate that the initial blue nodes must be more than half to ensure that the blue nodes ``dominate'' at the end, if one selects the initial blue nodes randomly. If the initial blue nodes are selected based on some smarter strategies such as centrality-based measures (such as degree, betweenness, and closeness), then a much smaller fraction of initial blue nodes is required for the blue color to be the winning color at the end.

Inspired by some prior work~\cite{Galam2007, Qian2015}, we also study the setup where the nodes have different levels of activeness and stubbornness; that is, some nodes are more probable to participate in the gatherings and some nodes are less inclined to follow the majority opinion. We then introduce some strategies for choosing initial seed nodes, which take these parameters into account, in addition to the graph structural properties.

The paper is organized as follows. We first give some basic definitions and an overview of some prior work in the rest of this Section. Our findings, which address questions (1) and (2), are presented in Sections~\ref{sec-conv} and~\ref{sec-inf}, respectively.

\subsection{Preliminaries}
\subsubsection{Graph Definitions}
\label{sec-Gdef}

\begin{itemize}
  
  \item{Graph}
 
  Let $G=(V, E)$ be a simple connected undirected graph. Let $n := |V|$ and $m := |E|$ represent the number of nodes and the number of edges, respectively. For a given node, $v \in V$, $N(v) := \{u \in V : \{u, v\} \in E\}$ represents the neighborhood of $v$. Furthermore, let $d(v) :=|N(v)|$ represent the degree of node $v$.

  \item{Expander\;Graph}

  An expander graph is expected a graph with good connectivity. There are three parameters often being used to measure the expansion of a graph: vertex expansion, edge expansion, and spectral expansion. A graph will be called here an expander graph if it has strong expansion properties. 

  \begin{enumerate}
 
    \item[(1)] {Vertex Expansion.} Let $S\subseteq V$ be a subset of nodes, and $\overline{S} := V \symbol{92} S$. 
    Let 
    \linebreak  
    $\partial S_{\rm out} := \{v \in \overline{S} | \exists u \in N(v),\, \text{such that} \: u \in S \}$, which is the outer boundary of set $S$. The vertex expansion is defined as $h_{\rm out}(G) 
    := \min_{\substack{0<|S|<\frac{n}{2}}} {|\partial S_{\rm out}|}/{|S|}$. For nodes in any ``small'' subset $S$ (with a size less than ${n}/{2}$) 
    of $V$, the greater the $h_{\rm out}(G)$ is, the larger the number of their neighbors outside $S$ is, and the better connected the graph $G$ is. Therefore, a graph $G$ with a greater $h_{\rm out}(G)$ has stronger expansion properties.

    \item[(2)] {Edge\;Expansion.} Let $S\subseteq V$ be a subset of nodes, and $\overline{S} := V \symbol{92} S$. Let \linebreak  $\partial S := \{\{u, v\} \in E | u \in S, v \in \overline{S} \}$, which is the boundary of set $S$. The edge expansion is defined as 
    $h(G) := 
    \min_{\substack{0<|S|<\frac{n}{2}}}
    {|\partial S|}/{|S|}$. For the number of edges from the nodes inside any ``small'' subset (with a size less than ${n}/{2}$) of $V$ to the nodes outside this subset, the greater the $h(G)$ is, the larger the number of edges is, and the better connected the graph $G$ is. Therefore, a graph $G$ with a greater $h(G)$ has stronger expansion properties.

    \item[(3)] {Spectral\;Expansion.} Let $A$ be the adjacency matrix of the graph $G$. Since $A$ is symmetric and real, it has $n$ real-valued eigenvalues. Let $\lambda (G)$ represent the second-largest absolute eigenvalue of $A$. A graph $G$ with a smaller value of $\lambda (G)$ has stronger expansion properties.
   
 \end{enumerate}

\end{itemize}

\subsubsection{Model}

\begin{figure}[H]

  \centering
  \includegraphics[width=0.6\textwidth]{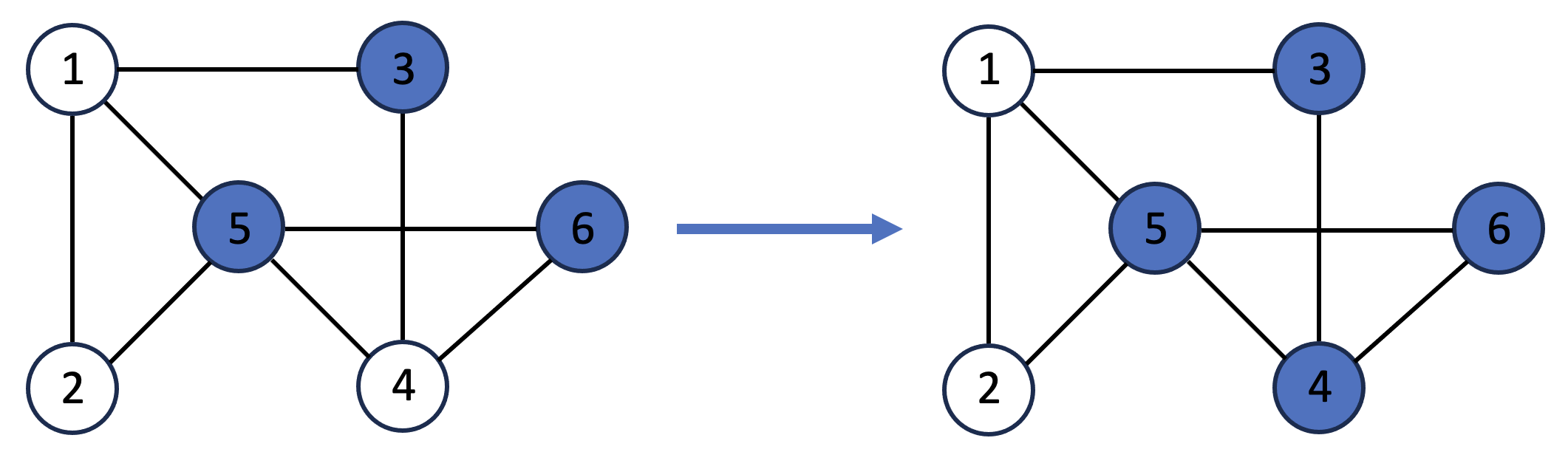}
  \caption{An example of one round of the Graph-based Galam model: the initial state of the graph ({\bf left}) and the updated graph after one round ({\bf right}).
  \label{fig1}}
\end{figure}

\begin{itemize}

  \item{Majority\;Model\;and\;Random\;Majority\;Model}

  Consider a graph $G$ that represents a social network. Each node in the graph is either blue or white at the initial state, which represents a person holding a positive or negative opinion about a product or a topic. In each round, all the nodes simultaneously update their color to the most frequent color among their neighbors. If there is a tie, the node keeps its color. This is known as the Majority Model. The Random Majority Model is the same as the Majority Model, except for the tie-breaking rule. In the Random Majority Model, a node chooses blue or white with an equal probability of 0.5 in case of a tie.

  \item{Galam\;Model}
  
  Consider a population of $n$ individuals (nodes) who can hold a positive or negative opinion on a subject (i.e., are blue or white). Furthermore, consider a set of rooms of various sizes, such that the summation of the size of the rooms is equal to $n$. In each round, all individuals are randomly assigned to these rooms. Then, all individuals simultaneously update their opinion to the most frequent opinion in the room. If there is a tie, then a tie-breaking rule is applied to handle the situation. In the original description of the model, individuals choose negative in case of a tie~\cite{Galam2002}. However, other variants are considered, for example, random tie-breaking rules~\cite{Galam2005}.

  \item{Graph-based\;Galam\;Model}

  In the present paper, we introduce a graph-based generalization of the Galam model, GGM. A graph $G=(V, E)$ is used to represent the society to be considered. The nodes and edges in the graph represents the individuals and the relationship between them, respectively. We define a coloring function \(\mathcal{C} \): $V \rightarrow \{w, b\}$, where $w$ and $b$ represent white and blue, respectively, which correspond to negative and positive.

  There are two steps in each round of this model: (1) randomly assign all nodes into groups with different sizes; and (2) update the color of each node following the local majority-based rule in the group.

  \begin{enumerate}
    \item[(1)]{Group\;Allocation.} All the cliques of size 1 to $R$ in the graph are collected and stored in a list. In our setup, we use $R=3$, but the model is well defined for larger values of $R$. Each time, one clique in the list is randomly picked with an equal likelihood. For each node in the clique that is picked, the remaining cliques that contain the node is removed from the clique list. This continues until the nodes are partitioned into cliques of size 1, 2, and 3 (the clique list is empty).

    \item[(2)] {Color\;Updating.} All the nodes simultaneously update their color to the most frequent color in their group. If there is a tie, a binary random number (0 or 1) is generated with equal probability. The nodes becomes blue if the random number is 1 and white otherwise.
  \end{enumerate}

  An example is given in Figure~\ref{fig1}. The graph on the left-hand side is the initial state. The list of all cliques of this graph is \{\{1\}, \{2\}, \{3\}, \{4\}, \{5\}, \{6\}, \{1, 2\}, \{1, 3\}, \{1, 5\}, \{2, 5\}, \{3, 4\}, \{4, 5\}, \{4, 6\}, \{5, 6\}, \{1, 2, 5\}, and \{4, 5, 6\}\}. One possible outcome of group allocation is \{\{1, 2\}, \{3\}, and \{4, 5, 6\}\}. In this case, node 1 and node 2 have the same color and keep their color. Node 3 is in a group of itself and it also keeps the color. Nodes 4, 5, and 6 are in the same group, and the major color in this group is blue. Hence, node 4 updates its color from white to blue. The state of the graph after this round is illustrated in the graph in Figure~\ref{fig1}, right.
\end{itemize}

\subsection{Prior Studies}
\label{Prior}

\begin{itemize}

  \item {Models}

  Numerous models that simulate the spreading of information and the formation of opinions have been introduced and studied, such as the Independent Cascade (IC) model~\cite{Kempe2003}, the Linear Threshold (LT) model~\cite{Kempe2003, Talukder2019}, and majority-based models~\cite{Zhuang2020, Avin2019, Amir2023, Anagnostopoulos2021}. 

  The Galam model is one of the most studied models in the area of sociophysics. The model was originally designed to explain how an initial minority can finally win the debate~\cite{Galam2002}. This model has become an established model in sociophysics and various variants of this model have been investigated. Some of these variants involve considering three different opinions~\cite{Gekle2005}, changing the biased tie-breaking rule to a random one~\cite{Galam2005}, adding inflexible individuals~\cite{Galam2007}, and introducing the level of activeness~\cite{Qian2015}. The model studied in this paper falls under the umbrella of this line of models.

  \item{Convergence\;time}

  Svatopluk Poljak and Daniel Turz\'{i}k proved that the convergence time of the majority model is upper bounded by a linear function of the number of edges; that is, $\mathcal{O}(m)$~\cite{Poljak1986}. Furthermore, some stronger bounds have been proven for some special types of graphs; see~\cite{Zehmakan2023}. Studies on convergence property have also been conducted for other majority-based models~\cite{Abdullah2020, Cruise2014}. For the random majority model on a cycle graph, the convergence time is proven to be in $\mathcal{O}(n^2)$~\cite{Zehmakan2023}. For the classic Galam model, Bernd G\"{a}rtner and one of the authors of this paper proved that the convergence time is in $\mathcal{O}(\log\log n)$ when all groups in the model have a size greater or equal to 3~\cite{Gartner2015}.

  \item{Influential\;nodes}
  
  The research about viral marketing in social networks has become popular in recent decades. A small set of seed individuals who hold a positive (blue) opinion on some subject needs to be selected to maximize the final number of individuals holding a positive opinion in the social network. Such a problem about how to select seed individuals is typically known as influence maximization (IM). David Kempe, Jon Kleinberg and \'Eva Tardos provided a foundation for this problem in their seminal paper in 2003~\cite{Kempe2003}. After that, there has been extensive study on the IM problem~\cite{Talukder2019, Auletta2020, Karia2022, Schoenebeck2020}. 

  The above IM problem can be modeled as a discrete optimization problem and it is usually proven to be computationally hard to solve for various models; more precisely, it is known to be non-deterministic polynomial-time hard (NP-hard); see~\cite{Kempe2003, Schoenebeck2020}. Therefore, several approximation, randomized, and heuristic algorithms, such as centrality-based algorithms~\cite{Kundu2011, Chen2009} (which choose nodes with the highest degree, betweenness, closeness, or pagerank) and greedy approaches~\cite{Leskovec2007, Goyal2011, Wu2015}, have been proposed. In recent years, machine learning techniques have become popular, which leads to the development of some machine learning-based algorithms for the IM problem~\cite{Kamarthi2020, Zhao2020, Li2023}.

  For the majority model, it was proven that the minimum size of initial blue nodes that leads to all nodes becoming blue in a random $d-$regular graph is $n/2$~\cite{Gartner2018}. Furthermore, the study of the majority model on graphs from real-world social networks showed that the nodes with the highest degrees have a very significant 'influence factor' than most other nodes~\cite{Out2021}.
  
\end{itemize}

\subsection{Experimental Setup}

All graph data for the experiments are from Stanford Network Analysis Project (SNAP)~\cite{url-SNAP}. Experiments were conducted on Facebook (FB) and Twitch Spain (T-ES)'s social networks. Some basic properties of these graphs are listed in Table~\ref{tab1}. Furthermore, several experiments focus on cycle graphs and complete graphs to examine the relationship between the number of nodes and the convergence time.

\begin{table}[H]
  \small
  \caption{Basic properties of the two examined social networks.
   Here, $n$, $m$ and ''AvgDegree'' denote the number of nodes, the number of edges and the average degree, respectively.}
  \label{tab1}
  \centering
  \begin{tabular}{c c c c}
    \toprule
    \textbf{Social Network Name} \hspace{1.5cm}	& $\boldsymbol{n}$ \hspace{2cm}	& $\boldsymbol{m}$ \hspace{1.5cm} &\textbf{Avg. Degree}\\
    \midrule
     Facebook \hspace{1.5cm}  & 4039 \hspace{2cm}	    &88,234 
     \hspace{1.5cm}     & 43.69\\

     Twitch Spain \hspace{1.5cm}       & 4638 \hspace{2cm}		& 59,382 \hspace{1.5cm}    & 25.55\\
    \bottomrule
  \end{tabular}
\end{table}

Since our model is a random process, all experiments were executed 20 times, {unless otherwise specified}, and the average output was considered. All experiments were conducted on an Apple M1 Pro chip, with 32~GB RAM and MacOS Ventura ({\textit{Apple\, Inc.},  
Cupertino, CA, USA}).


\section{Convergence Time}
\label{sec-conv}

Some experiments were conducted to study the convergence time of the GGM in cycle graphs, complete graphs, and some other types of special graphs in this Section. Before discussing the outcomes of these experiments, we show that the GGM always converges on a connected graph.

\textbf {Convergence to an Absorbing State.} For a connected graph $G$ with $n$ nodes, the model on such a graph can be regarded as a Markov chain. Since there are $2^n$ possible colorings in the model, this Markov chain has $2^n$ states. If the probability of transition from state $s$ to $s'$ is non-zero, then there is a directed edge from $s$ to $s'$. A state is called an absorbing state if there is no edge going out, and a Markov chain is called an absorbing Markov chain if every state in the Markov chain can reach an absorbing state.

In GGM, the two states where all nodes have the same color (blue or white) are two absorbing states. Since each node in the graph keeps its color in such two states, there is no outgoing edge from the two states. Except these two states, any other state has at least two edges going out from it and there is a path from the state to the absorbing states. Consider a state that there are $i$ blue nodes ($i\neq 0$ and $i\neq n$, i.e., there are not $n$ blue nodes or $n$ white nodes). Since the graph is connected, there must be at least two adjacent nodes with different colors. In the coming round, there is a non-zero probability that these two nodes are assigned to the same group and all other nodes are assigned to other groups of size one. Then, these two 
nodes become both blue or white and all other nodes keep their color at the end of this round. Therefore, the state that has $i$ blue nodes must have at least two outgoing edges: one edge goes to a state that has $(i+1)$ blue nodes and the other one goes to a state that has $(i-1)$ blue nodes. As the process keeps going, the state of $i$ blue 
nodes finally reach the state that all the nodes are blue (or white). Hence, the GGM on a connected graph is an absorbing Markov chain with two absorbing states where all nodes are blue or white. Furthermore, the process eventually converges to one of the absorbing states.

\subsection{Cycle Graph}

As mentioned in Section~\ref{Prior}, prior studies have shown that the convergence time of the random majority model on a cycle graph with $n$ nodes is in $\mathcal{O}(n^2)$. Since one can think of the GGM somewhat as a room-based random majority-based model, let us speculate that the convergence time of the model in a cycle graph with $n$ nodes is also of order $\mathcal{O}(n^2)$ for GGM.

\begin{figure}[H]
     \hfill 
     \centering
     \begin{subfigure}[b]{0.2\textwidth}
         \centering
         \includegraphics[width=\textwidth]{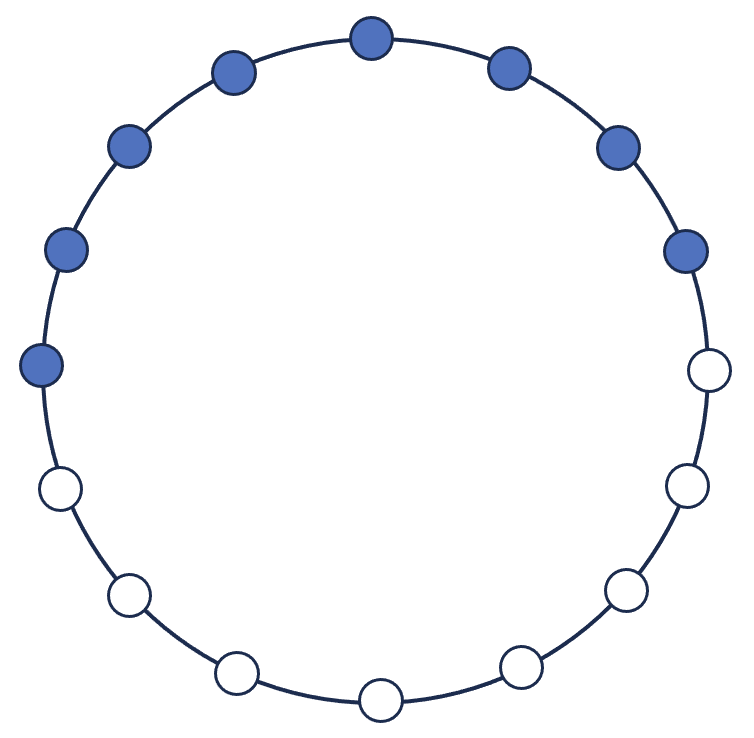}
         \captionsetup{justification=centering}
         \caption{}
         \label{fig2.1}
     \end{subfigure}
     \hfill
     \begin{subfigure}[b]{0.2\textwidth}
         \centering
         \includegraphics[width=\textwidth]{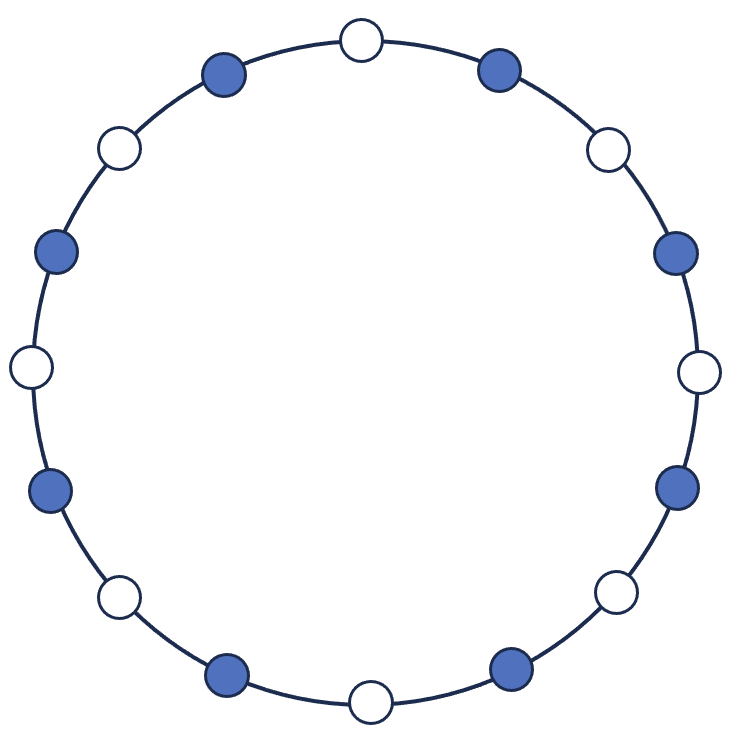}
         \captionsetup{justification=centering}
         \caption{}
         \label{fig2.2}
     \end{subfigure}
     \hfill
    \captionsetup{justification=justified}
    \caption{An example showing the initial state of a cycle: (\textbf{a}) all blue nodes are next to each other; (\textbf{b}) white and blue nodes are alternating in the cycle.}\label{fig2}
\end{figure}

The experiments were conducted on 10 cycle graphs with the number of nodes selected from 100 to 1000 every 100. The number of initial blue nodes and how to set these blue nodes should be determined to maximize the convergence time. One can expect that it would take more time for the GGM to converge when the number of blue and white nodes are equal. If blue nodes are more (or less) than white nodes, the process tends to converge to a state of all nodes being blue (or white) more quickly. Therefore, the initial number of blue nodes were set to be half of the total nodes. In addition, the blue nodes were selected in two ways. The first way is that all the blue nodes are next to each other to form a path. The other way is that all blue nodes and white nodes are alternating in the cycle. An example of these two types of initial state can be seen in Figure~\ref{fig2}. For the graph in Figure~\ref{fig2}a, only four nodes have a non-zero probability to change their color in the coming round. However, all the nodes in the graph shown in Figure~\ref{fig2}b have a non-zero probability to change their color. All the other configurations of initial blue nodes 
are expected to place between these two special cases, with regard to convergence time.

\par\medskip

\begin{figure}[H]
    \centering
    \includegraphics[width=0.6\textwidth]{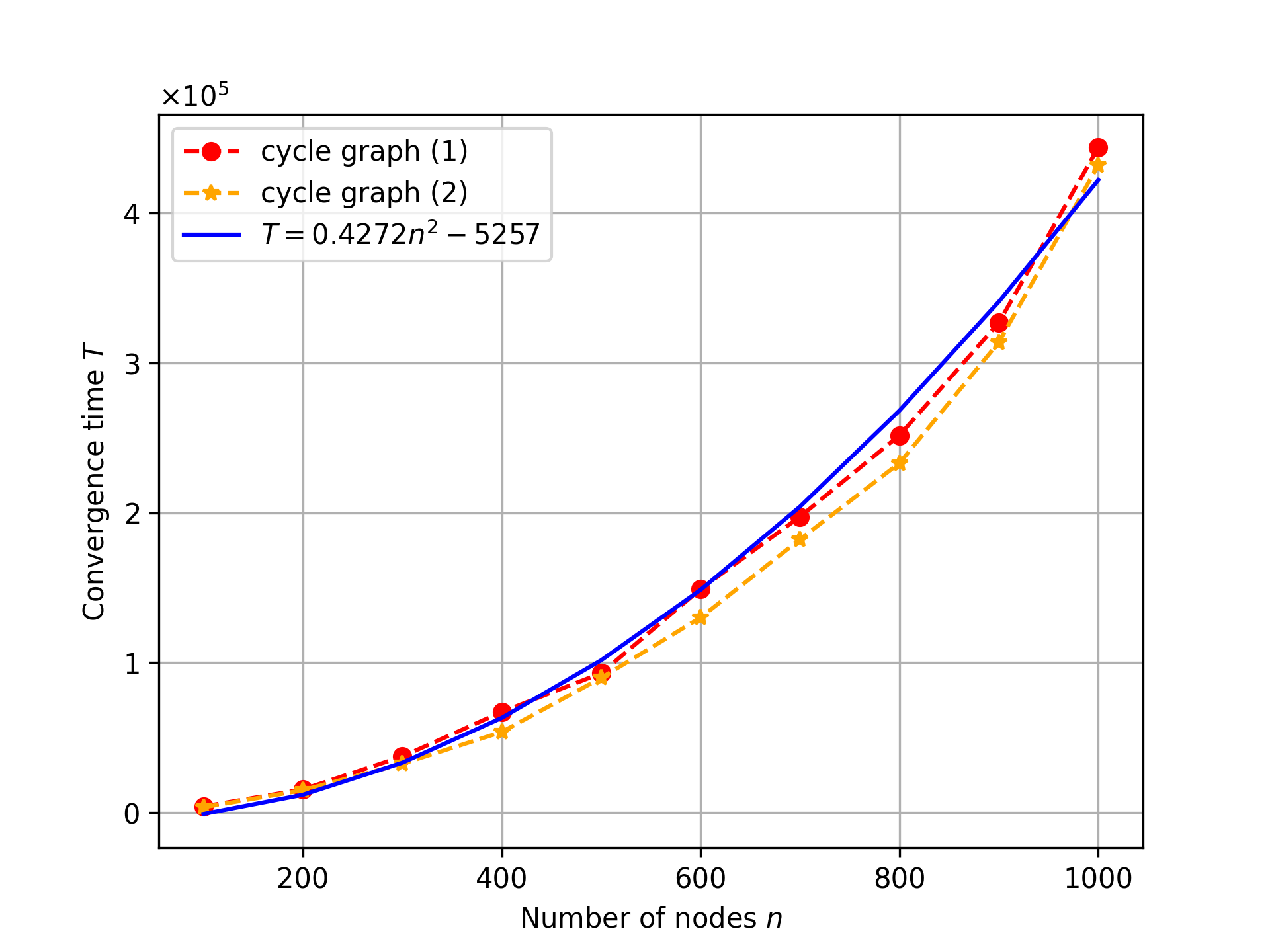}
    \captionsetup{justification=justified}
    \caption{Convergence time in cycle graph\label{fig3}}
\end{figure}

Figure~\ref{fig3} shows the outcome of the experiments. Cycle graph (1) and cycle graph (2) correspond to the two ways of initializing. The experiment result indicates that the convergence time of the GGM on cycle graphs is of order $\mathcal{O}(n^2)$. When half of the nodes are blue, the two different ways of setting these blue nodes have little impact on the convergence time.

\unskip
\hfill

\subsection{Complete Graph}
Prior study \cite{Gartner2015} has shown that the convergence time of the original Galam model with all groups of a size greater or equal to 3 is of order $\mathcal{O}(\log\log n)$. In the original Galam model, any two individuals in the social network can be assigned in the same group, which means that there exists a relationship between them. If a graph is used to represent the social network, there must be an edge between any two nodes in this graph. Then, the original Galam model can be regarded as a complete graph. Therefore, one might speculate that the convergence time of the GGM on complete graphs with $n$ nodes is of order $\mathcal{O}(\log\log n)$.

The experiment was conducted on 10 complete graphs with the number of nodes selected in the set \{50, 100, 200, 500, 1000, 2000, 3000, 4000, 5000, and 10,000\}. At the initial state, half of the nodes were blue, since as expected, this would result in the largest convergence time. Since each node in a complete graph is connected to all other nodes and there is no structural difference between the nodes, the blue nodes were just randomly selected. 

\begin{figure}[H]
     \centering
     \hfill
     \begin{subfigure}[b]{0.49\textwidth}
         \centering
         \includegraphics[width=\textwidth]{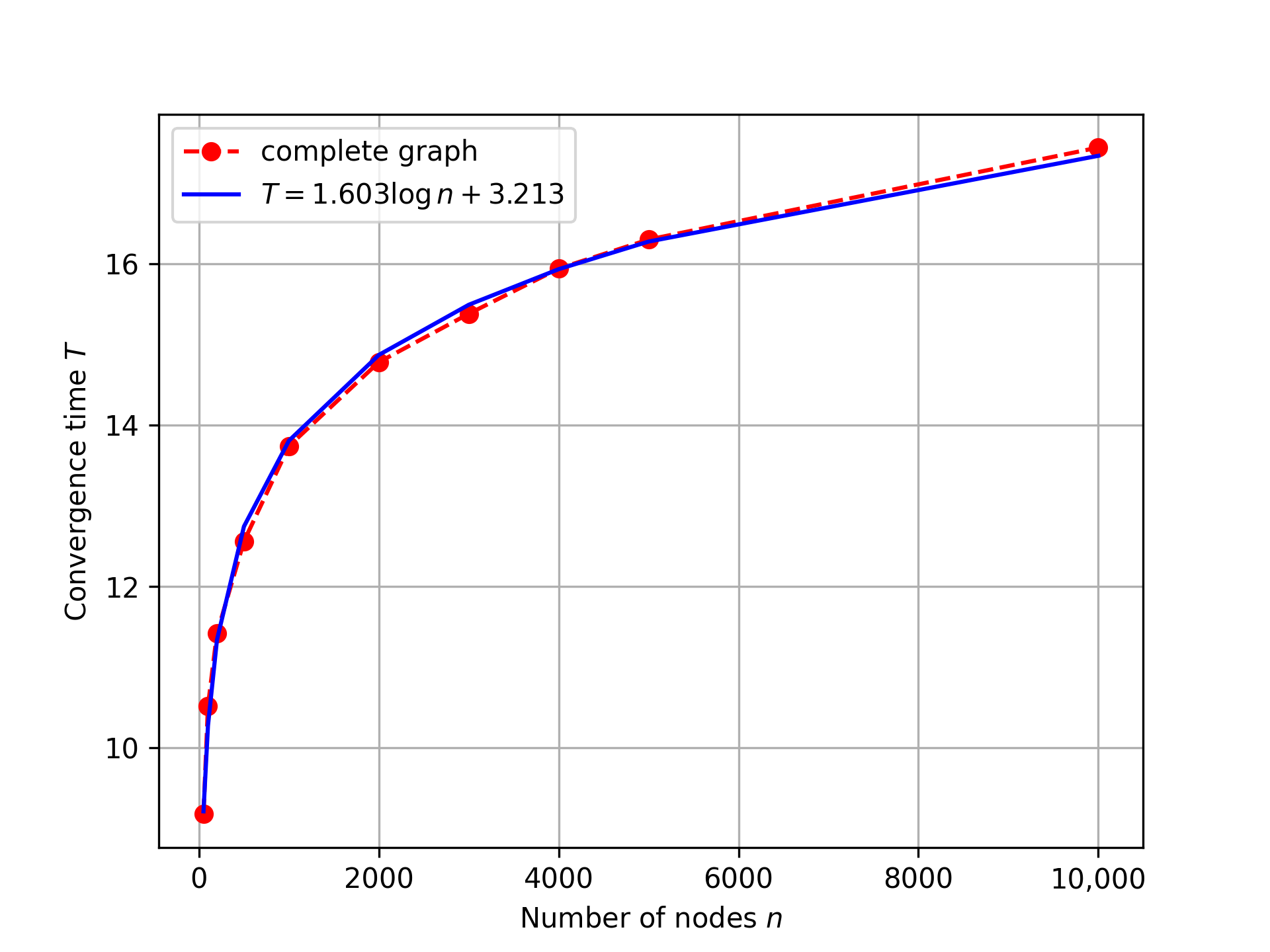}
         \captionsetup{justification=centering}
         \caption{}
         \label{fig4.1}
     \end{subfigure}
     \hfill
     \begin{subfigure}[b]{0.49\textwidth}
         \centering
         \includegraphics[width=\textwidth]{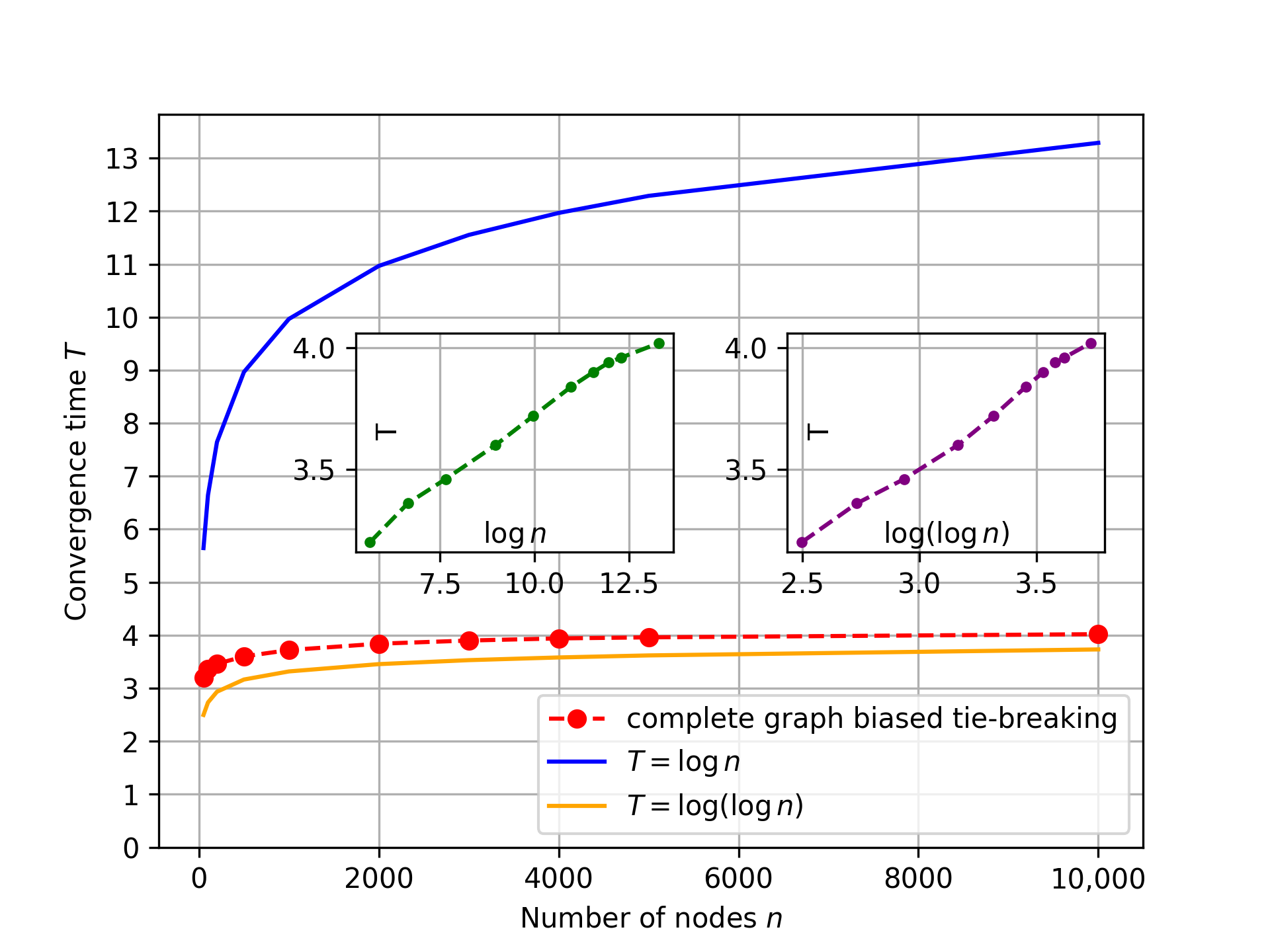}
         \captionsetup{justification=centering}
         \caption{}
         \label{fig4.2}
     \end{subfigure}
     \hfill
    \captionsetup{justification=justified}
    \caption{Convergence time in complete graph with (\textbf{a}) random tie-breaking rule and (\textbf{b}) biased tie-breaking rule.}
    \label{fig4}
\end{figure}

The outcome of the experiment is depicted in Figure~\ref{fig4}a. It should be noted that the convergence time of the GGM on complete graphs is of order $\mathcal{O}(\log n)$, but not $\mathcal{O}(\log\log n)$. The main reason for the difference is more expected to be the tie-breaking rule and the number of groups of even size. In the original Galam model, all the people in the group hold a negative (white) opinion if there is a tie. It is a biased tie-breaking rule, which can lead to fast convergence if there are more groups with a size of an even number. In addition, the convergence time $\mathcal{O}(\log\log n)$ has been proved when all groups have a size of equal to or greater than 3. However, the group size in the GGM in the present paper is less than or equal to 3, which might slow down the diffusion of opinions. 

\par\medskip

We conducted another experiment on those 10 complete graphs with the biased tie-breaking rule. The maximum size of a group was increased from 3 to 4. Figure~\ref{fig4.2} shows the result of this experiment. There are two insets in Figure~\ref{fig4.2}, which illustrate the plots of the convergence time, $T$, against $\log n$ and $\log (\log n)$. The insets show that the convergence time is a linear function of both the logarithm and double logarithm of $n$. The reason is that $\log (\log n)$ is a linear function of $\log n$ when $n$ is not big enough. Thus, experiments should be executed for $n$ that is large enough to determine the order of the convergence time. However, we cannot run the experiments for 
so large $n$ due to the limitation of computing power. In such a case, we plot the curves of the convergence time $T$ against the number of nodes $n$, $T = \log (\log n)$ and $T = \log n$ in Figure~\ref{fig4.2} to make a direct comparison. It can be observed that the curve of $T$ against $n$ is much closer to the curve of $T = \log (\log n)$. It indicates that the convergence time of the GGM on a complete graph looks to be rather of order $\mathcal{O}(\log\log n)$ if the biased tie-breaking rule is followed and the maximum group size is increased and is an even number.

\subsection{Some Other Special Graphs}

The outcomes of the experiments on the cycle and complete graphs from the previous sections show that the GGM converges much faster on a complete graph than on a cycle graph if they have the same $n$. For an $n$-node graph, the complete graph 
has $\binom{n}{2} = {n(n-1)}/{2}$ edges, but a cycle graph only has $n$ edges. One may expect that adding more edges to a graph can increase the convergence rate. 

\begin{figure}[H]
     \hfill
     \centering
     \begin{subfigure}[b]{0.16\textwidth}
         \centering
         \includegraphics[width=\textwidth]{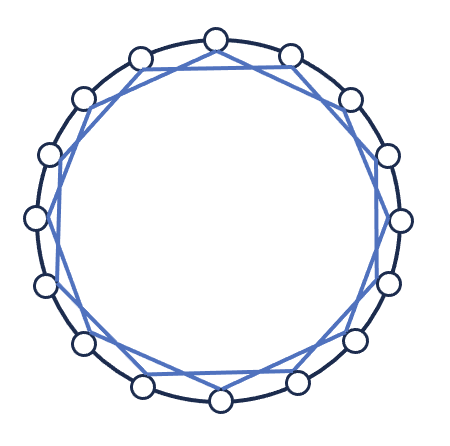}
         \captionsetup{justification=centering}
         \caption{}
         \label{fig5.1}
     \end{subfigure}
     \hfill
     \begin{subfigure}[b]{0.16\textwidth}
         \centering
         \includegraphics[width=\textwidth]{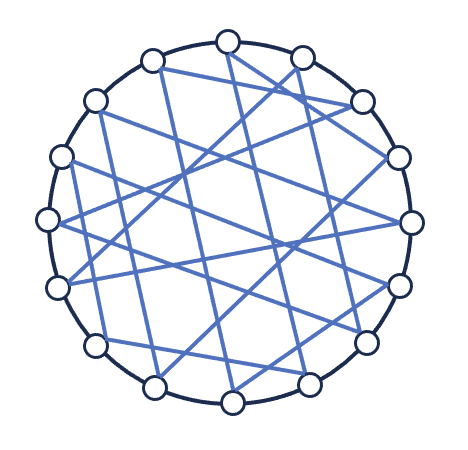}
         \captionsetup{justification=centering}
         \caption{}
         \label{fig5.2}
     \end{subfigure}
     \hfill
       \begin{subfigure}[b]{0.25\textwidth}
         \centering
         \includegraphics[width=\textwidth]{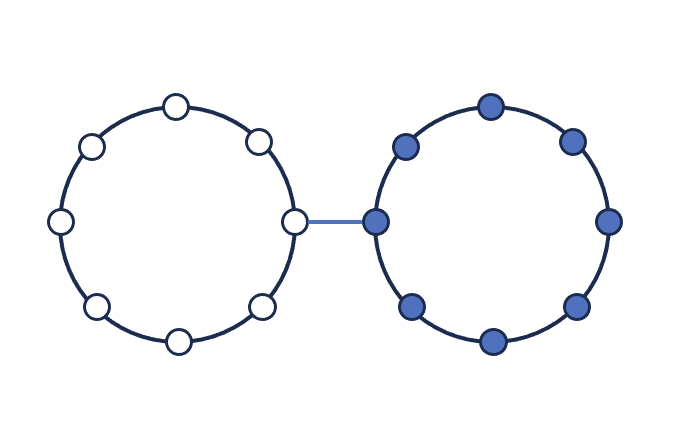}
         \captionsetup{justification=centering}
         \caption{}
         \label{fig5.3}
     \end{subfigure}
     \hfill
       \begin{subfigure}[b]{0.25\textwidth}
         \centering
         \includegraphics[width=\textwidth]{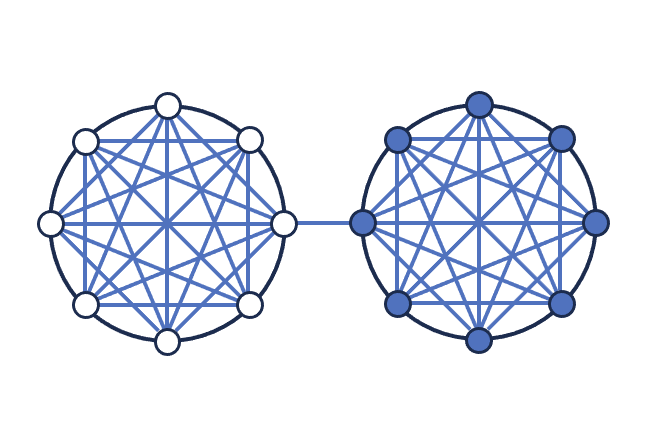}
         \captionsetup{justification=centering}
         \caption{}
         \label{fig5.4}
     \end{subfigure}
     \hfill
    \captionsetup{justification=justified}
    \caption{(\textbf{a}) Two-cycle, (\textbf{b}) random cycle, (\textbf{c}) two-cycle-connected and (\textbf{d}) two-complete-connected graphs.}
    \label{fig5}
\end{figure}

An experiment was executed on two-cycle graphs (Figure~\ref{fig5.1}) and random cycle graphs (Figure~\ref{fig5.2}). A two-cycle graph can be generated by taking a cycle graph and adding an edge between every two nodes with a distance of 2. To build a random cycle graph, two randomly selected extra edges should be added to each node in a cycle graph. 

\par\medskip

\begin{figure}[H]
     \hfill 
     \centering
     \begin{subfigure}[t]{0.49\textwidth}
         \centering
         \includegraphics[width=\textwidth]{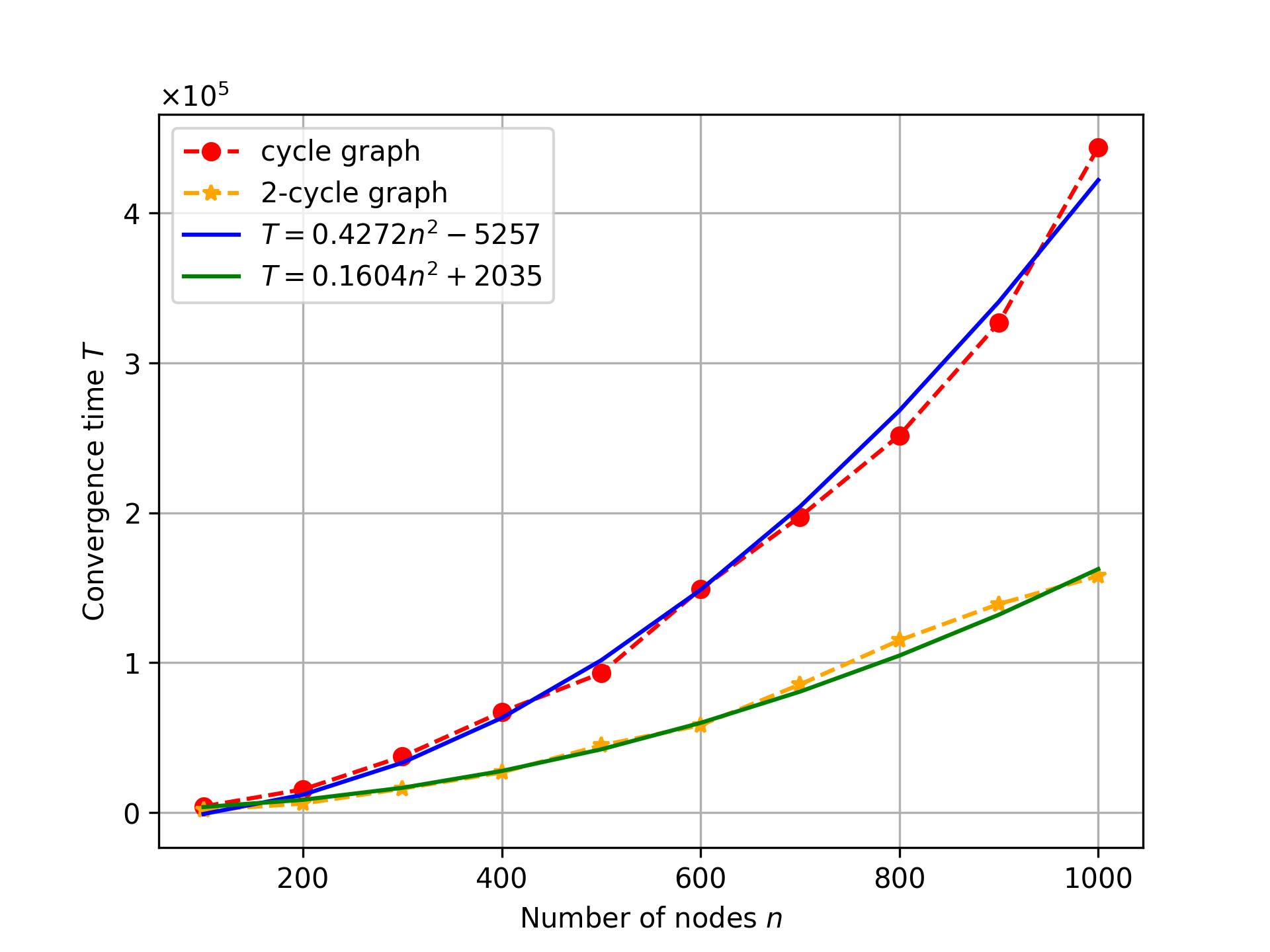}
         \captionsetup{justification=centering}
         \caption{}
         \label{fig6.1}
     \end{subfigure}
     \hfill
     \begin{subfigure}[t]{0.49\textwidth}
         \centering
         \includegraphics[width=\textwidth]{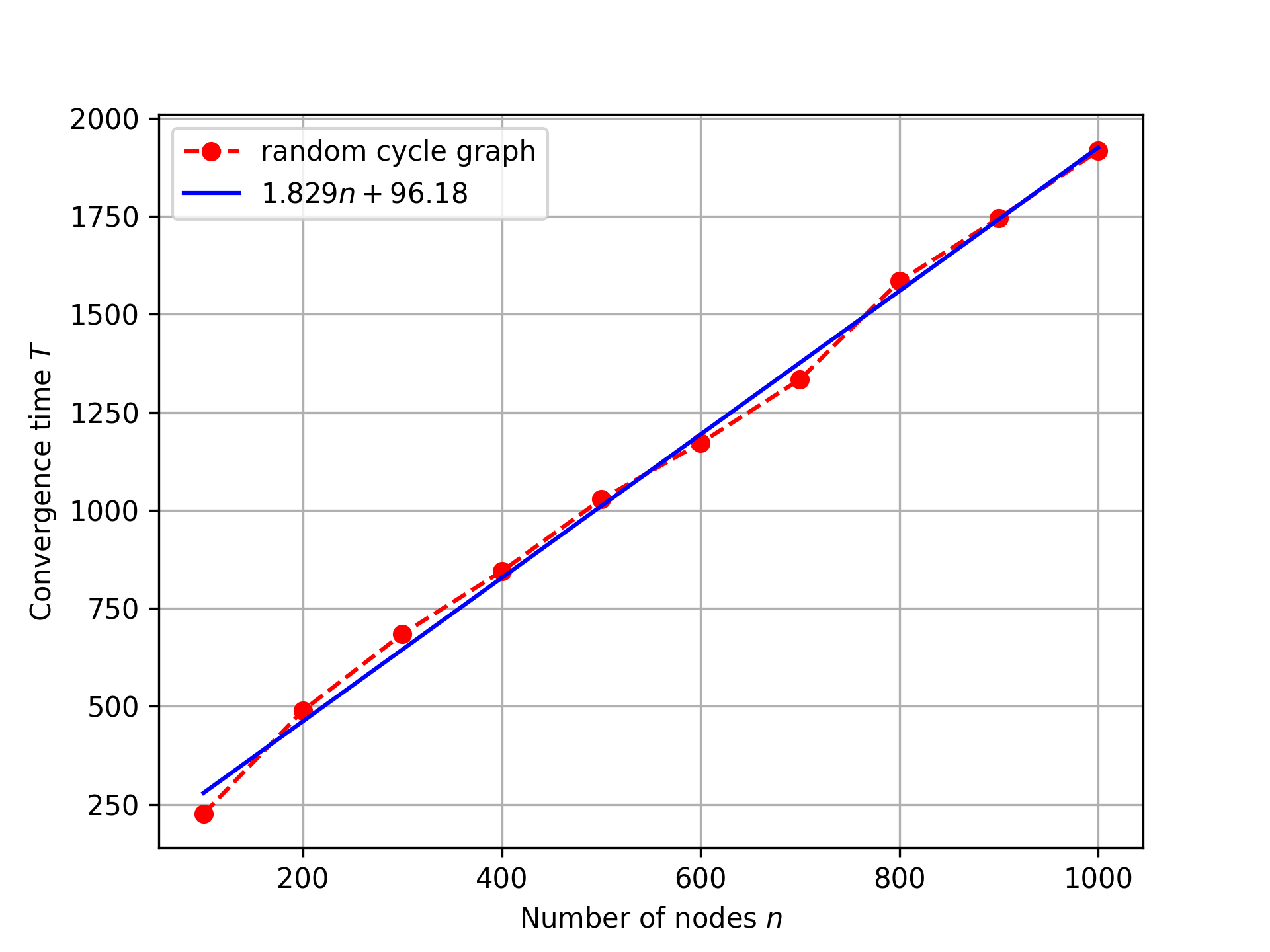}
         \captionsetup{justification=centering}
         \caption{}
         \label{fig6.2}
     \end{subfigure}
     \hfill
     \par\medskip
     \begin{subfigure}[t]{0.49\textwidth}
         \centering
         \includegraphics[width=\textwidth]{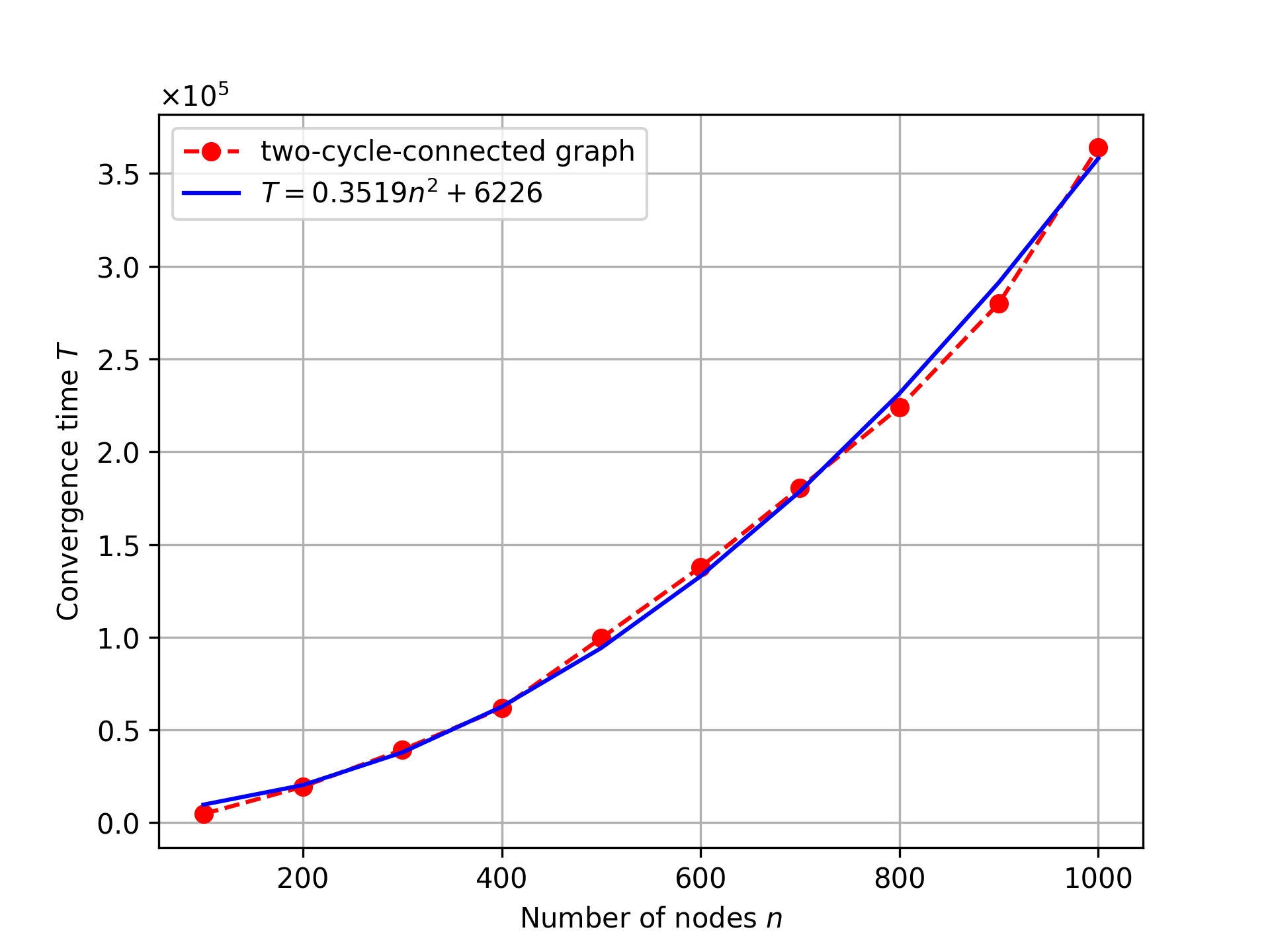}
         \captionsetup{justification=centering}
         \caption{}
         \label{fig6.3}
     \end{subfigure}
     \hfill
     \begin{subfigure}[t]{0.49\textwidth}
         \centering
         \includegraphics[width=\textwidth]{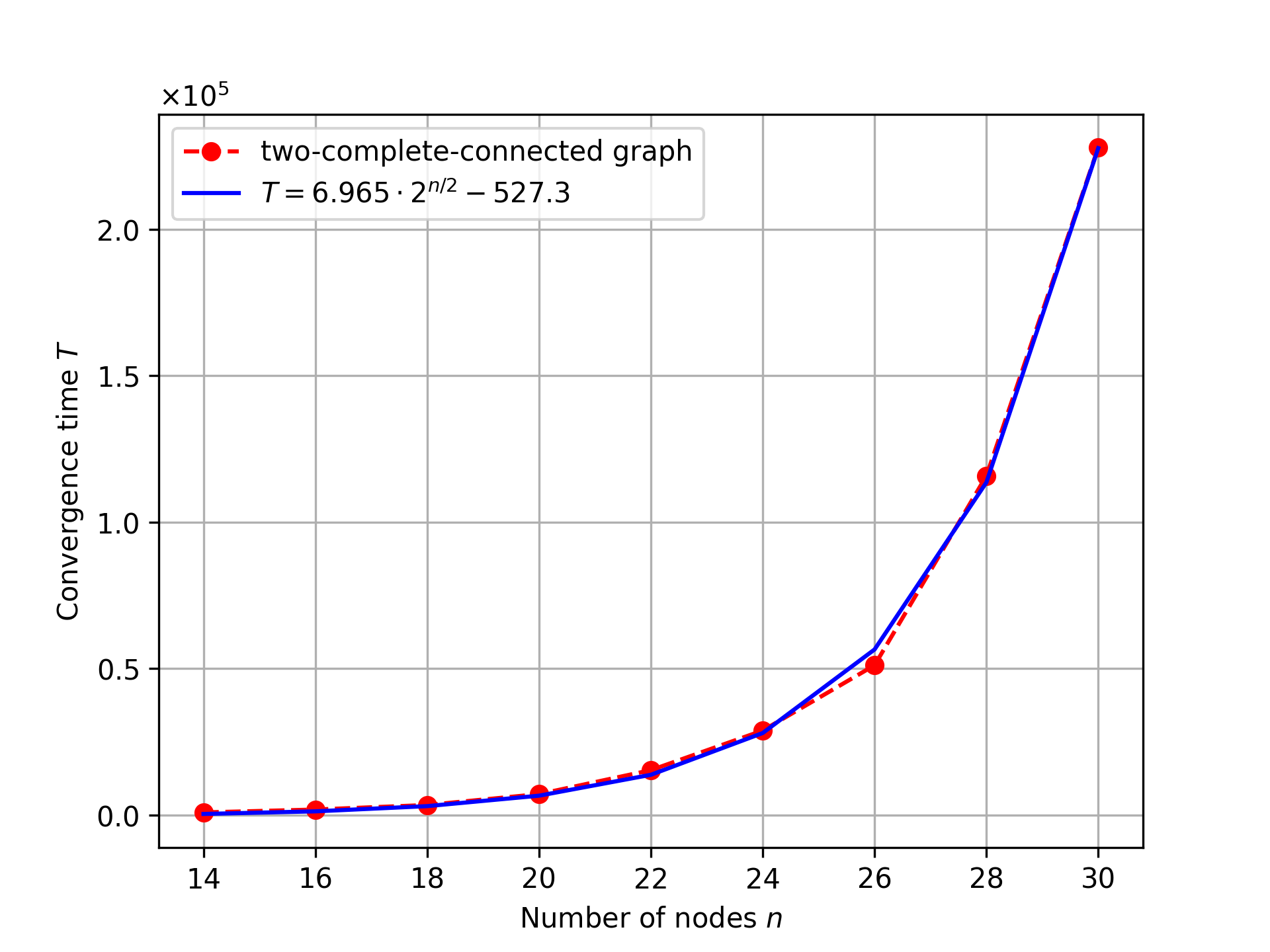}
         \captionsetup{justification=centering}
         \caption{}
         \label{fig6.4}
     \end{subfigure}
     \hfill
    \par\medskip
    \captionsetup{justification=justified}
    \caption{Convergence time in (\textbf{a}) two-cycle, (\textbf{b}) random cycle, (\textbf{c}) two-cycle-connected and (\textbf{d}) two-complete-connected graphs.}\label{fig6}
\end{figure}

Figure~\ref{fig6.1} and Figure~\ref{fig6.2} depicts the convergence time of the GGM on two-cycle graphs and random cycle graphs, respectively. It can be observed that the process converges faster on two-cycle graphs than on a cycle graph, but the convergence time is still of order $\mathcal{O}(n^2)$. However, the convergence time is of order $\mathcal{O}(n)$ on the random cycle graph. With the same $n$ and $m$ (number of edges), a two-cycle graph converges much slower than a random cycle graph. Therefore, the increase in the convergence speed is not merely the result of adding extra edges, but rather how to add these edges.

It should also be noted that adding extra edges improperly can even slow down the convergence of the process. Figure~\ref{fig5.3} and Figure~\ref{fig5.4} illustrate the 
two-cycle-connected (TCC) graph and the two-complete-connected (TKC) graph, respectively. These types of graphs can be built by connecting two cycle 
or complete graphs of the same size with one edge. It is obvious that the number of edges in a TKC graph is greater than that in a TCC graph if they 
have the same $n$. We run experiments on these two types of graphs. The outcome shows that the convergence time of the GGM on a TCC graph is 
$\mathcal{O}(n^2)$ (Figure~\ref{fig6.3}). However, the TKC graph has a convergence time of $\mathcal{O}(2^{n/2})$
(Figure~\ref{fig6.4}) even though it has more edges than a TCC graph with the same $n$. It takes much 
longer for a TKC graph to converge (i.e., more than 200,000 rounds for only 30 nodes).

\par\medskip

Considering the expansion property of TCC and TKC graphs, both of them have an edge expansion of $2/n$ and a node expansion of $2/n$. It means these two types of graphs have a poor connectivity (removing only one edge or one node can make them become disconnected). The spectral expansion of a TCC graph with 50 nodes is only 2.236 and it is much smaller than the spectral expansion of a 50-node TKC graph, which is 23.962. Therefore, the TCC graph has better expansion properties than the TKC graph.

Furthermore, in the case of a two-cycle and random cycle graph, while they have the same number of nodes and edges, the random cycle graph has much stronger expansion properties since the randomly added edges significantly improve the connectivity of the graph. Strong expansion and connectivity properties can increase the flow of information in the network. Consequently, the process is much faster on a random cycle graph than on a two-cycle graph, as observed in our experiments.

From the results of these experiments, it can be observed that the expansion property has an impact on the convergence time. For instance, a complete graph or random cycle graph is a good expander and the process in such a graph converges fast. One can expect that the convergence time of the GGM on a graph with strong expansion properties to be less. 


\section{Influential Nodes}
\label{sec-inf}

\begin{figure}[H]
    \centering
    \captionsetup{justification=justified}
    \includegraphics[width=0.6\textwidth]{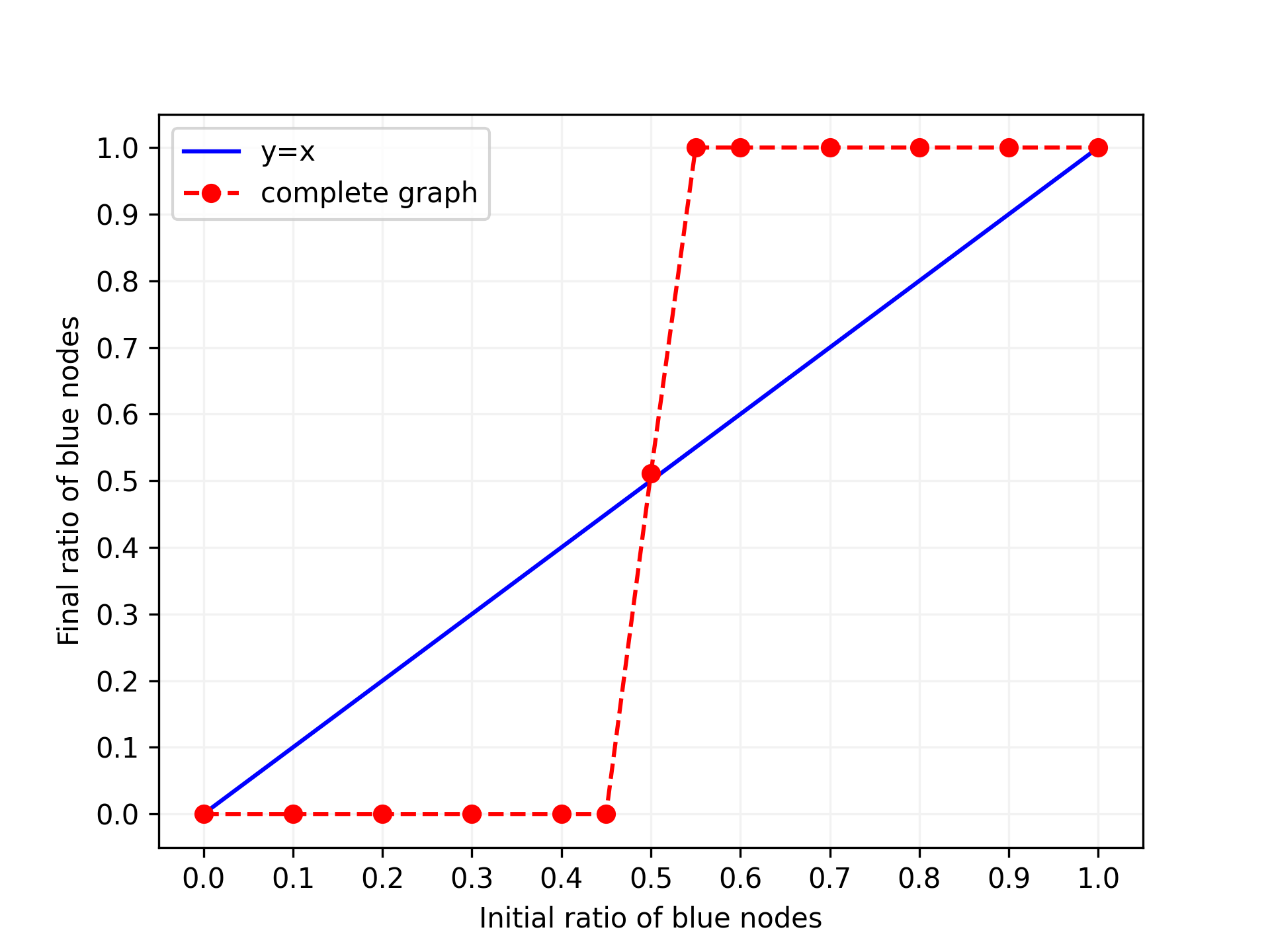}
    \caption{Randomly select initial blue nodes on a 1000-node complete graph.\label{fig7}}
\end{figure}

As discussed in Section \ref{sec-conv}, the original Galam model can be modeled by a complete graph. Since there is no difference between each node in a complete graph, only the number of initial blue nodes influences the total number of blue nodes at the end. Thus, the initial blue nodes are randomly selected. Experiments were conducted for the GGM 
on a 1000-node complete graph with different initial ratios of blue nodes selected in the set $\{0.1, 0.2, 0.3, 0.4, 0.45, 0.5, 0.55, 0.6, 0.7, 0.8, 0.9\}$. Experiments were executed 1000 times for each initial ratio of blue nodes. When the process converges to an absorbing state, the average final ratio of blue nodes is checked. Figure~\ref{fig7} shows 
that the GGM converges to the absorbing state that all nodes are blue (white) eventually if the initial ratio of blue nodes is just over (under) half. This looks to be true in any graph, which possesses some level symmetry (such as vertex-transitive graphs).

\par\medskip

However, most real-world social networks cannot be modeled as a complete graph. For instance, some individuals may have more connections to a social network than others. Such individuals are modeled as nodes with a higher degree. Selecting these nodes as initial blue nodes may lead to different results compared to selecting nodes with a lower degree. Therefore, a strategy should be developed to smartly choose the initial blue nodes rather than just randomly choosing them. One of the most often used approach is to use different centrality based measures.

\subsection{Centrality-Based Influential Nodes}

The centrality of a node can be used to identify the ``importance'' of the node~\cite{Newman2010}. There are many definitions of centrality, and the degree, betweenness, and closeness centrality are considered in the present paper.

Degree. As defined in Section~\ref{sec-Gdef}, it is the number of neighbors of a node: 
$$d(v):=|N(v)|.$$

Betweenness. If a node lies on as many shortest paths between other nodes in a graph, it is more probable that this node is placed 
in the center of the graph. Based on this idea, the betweenness centrality is defined as:
$$b(v) := \sum_{s\neq v\neq t}\frac{\sigma_{st}(v)}{\sigma_{st}}.$$
where $\sigma_{st}$ is the total number of shortest paths from node $s$ to node $t$, and $\sigma_{st}(v)$ is the number of those paths passing through the node $v$.

Closeness. If the total distance between a node and all other nodes is short, then the node is close to all other nodes. The closer a node to all other nodes, the more central the node is. The closeness centrality comes from this idea and it is defined as:
$$c(v) := \frac{n-1}{\sum_{u} d(u, v)}.$$
where $d(u, v)$ is the distance (shortest path) between node $u$ and node $v$.

To compare the difference between randomly selecting initial blue nodes and choosing initial blue nodes based on centrality, we 
created three lists for the graph on which our model ran. Each list stores all the nodes of the graph. Then, these lists were sorted in descending order of the degree, betweenness, and closeness centrality 
of the nodes, respectively. When setting the initial blue nodes, we choose the first $\lceil \alpha n \rceil$ ($\alpha$ is the ratio initial of blue nodes) nodes in the list as blue nodes.

We ran our model on the T-ES and FB social networks. Four experiments were designed on each social network. One experiment is for randomly selecting the initial blue nodes and the other three are for selecting the initial blue nodes according to the three types of centrality. It should be mentioned that there is often a time limit for a product to take over a market in the real world. We assumed that each person takes part in three discussions every day on average, and the time limit is 100 days. Therefore, we ran the GGM 300 rounds in each experiment instead of waiting for it to converge. Note that running the process until the final convergence can also be quite expensive computationally for these networks.

\begin{figure}[H]
     \centering
     \hfill
     \begin{subfigure}[b]{0.49\textwidth}
         \centering
         \includegraphics[width=\textwidth]{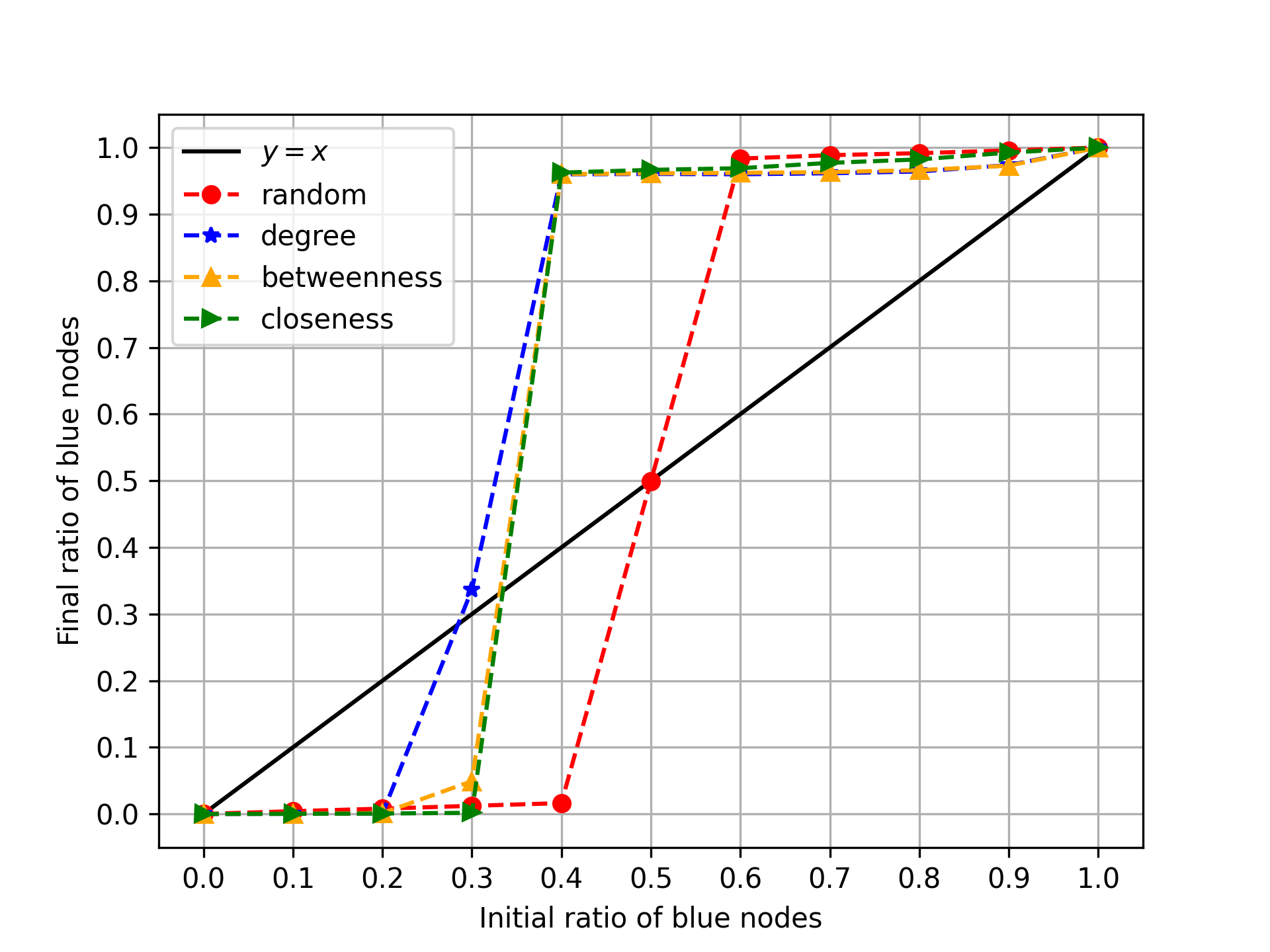}
         \captionsetup{justification=centering}
         \caption{}
         \label{fig8.1}
     \end{subfigure}
     \hfill
     \begin{subfigure}[b]{0.49\textwidth}
         \centering
         \includegraphics[width=\textwidth]{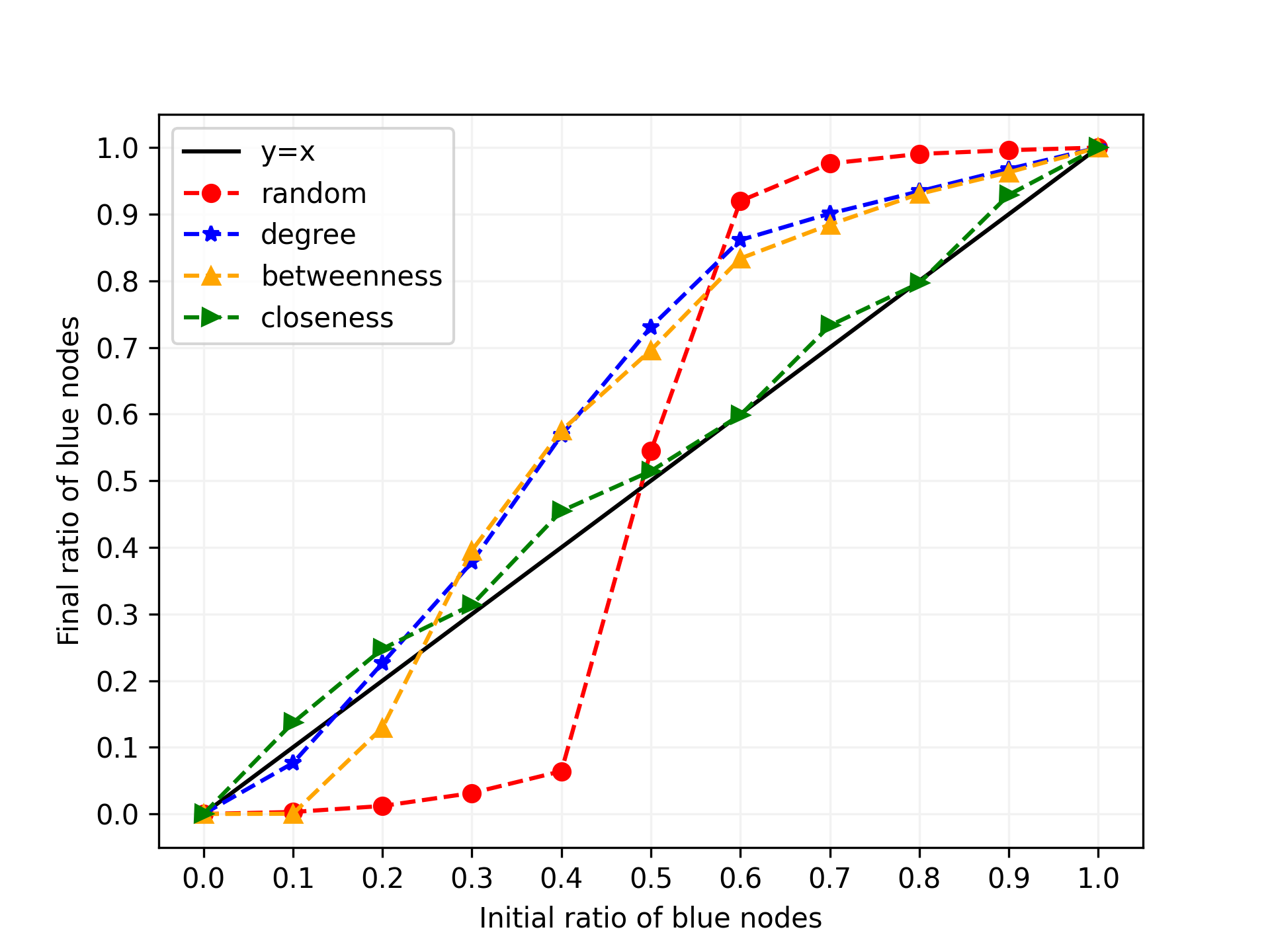}
         \captionsetup{justification=centering}
         \caption{}
         \label{fig8.2}
     \end{subfigure}
     \hfill
    \captionsetup{justification=justified}
    \caption{Centrality-based influential nodes selection on (\textbf{a}) Twitch Spain's and (\textbf{b}) Facebook's social networks.}
    \label{fig8}
\end{figure}

Figure~\ref{fig8} shows the outcomes of these experiments. When the initial blue nodes are randomly selected, the outcomes of the experiments on these two real-world social networks are quite similar to that on the complete graph, where 0.5 is a transition point. The final ratio is more than 0.9 (less than 0.1) and tends to be 1 (0) when the initial ratio is greater (less) than 0.5. On the other hand, if the initial blue nodes are selected according to the centrality-based measures, then even a smaller fraction of initial blue nodes can result in a final blue ratio that is greater than 0.5.

\par\medskip

For the T-ES social network, there is not much difference between choosing the initial blue nodes based on degree, betweenness, and closeness centrality. When the initial ratio is equal to or greater than 0.4, the final ratio is close to 1. For the FB social network, the difference between degree-based and betweenness-based strategies is not big. However, the initial and final ratios of blue nodes are almost the same when the initial blue nodes are selected according to their closeness. In fact, the number of blue nodes just oscillated around the initial number during the experiment. This may be attributed to the structure of FB's social network.

\subsection{Personality-Based Influential Nodes}

In addition to considering how ``central'' individuals in a social network are, one also needs to take into account their personality when picking up the initial blue nodes. For instance, some people are more active and they spend more time on social communication. These people are more expected to influence their connections than those who communicate little with others. In addition, some individuals are inflexible and they firmly stick to their vision. To maximize the final ratio of blue nodes, it might be a reasonable way to choose the initial blue nodes that are more active and inflexible.

In Refs.~\citep{Galam2007, Qian2015}, the ``inflexible agent'' and ``activeness'' have been introduced. In the present paper, we re-introduce these two parameters, called ``activeness'' and ``stubbornness'', for each node in a graph and define them in a slightly different way (which we believe is more self-explaining.) Each node is assigned a value between 0 and 1 for its activeness and a value between 0 and 1 for its stubbornness. In our experiments, two random numbers were generated from the uniform distribution (0, 1) independently and they were set as activeness and stubbornness, respectively. For a node, the greater the activeness (stubbornness) is, the more active (inflexible) is the node. More precisely, a node $v$ with activeness $act(v)$ decides to participate in the opinion exchange process with probability $act(v)$. Furthermore, for stubbornness $stub(v)$, it decides not to follow the majority 
updating rule and keep its opinion unchanged.

The GGM needs to be modified after these two parameters are introduced. Before each round of the model, a random number following uniform distribution (0, 1) is generated for each node in the graph. If the random number is greater than the activeness of a node, then the node is removed from the graph in this round. When each node is updating their colors, a random number following uniform distribution (0, 1) is generated for each group. If the stubbornness of a node in the group is greater than the random number, this node keeps its color, even though its color is not the majority in the group. 

Before experiments on the modified model were executed, three lists of all nodes were created. These three lists are sorted in descending order of the degree, activeness, and stubbornness, respectively. The first $\lceil \alpha n \rceil$ was selected as the initial blue nodes.

\begin{figure}[H]
     \centering
     \hfill
     \begin{subfigure}[b]{0.49\textwidth}
         \centering
         \includegraphics[width=\textwidth]{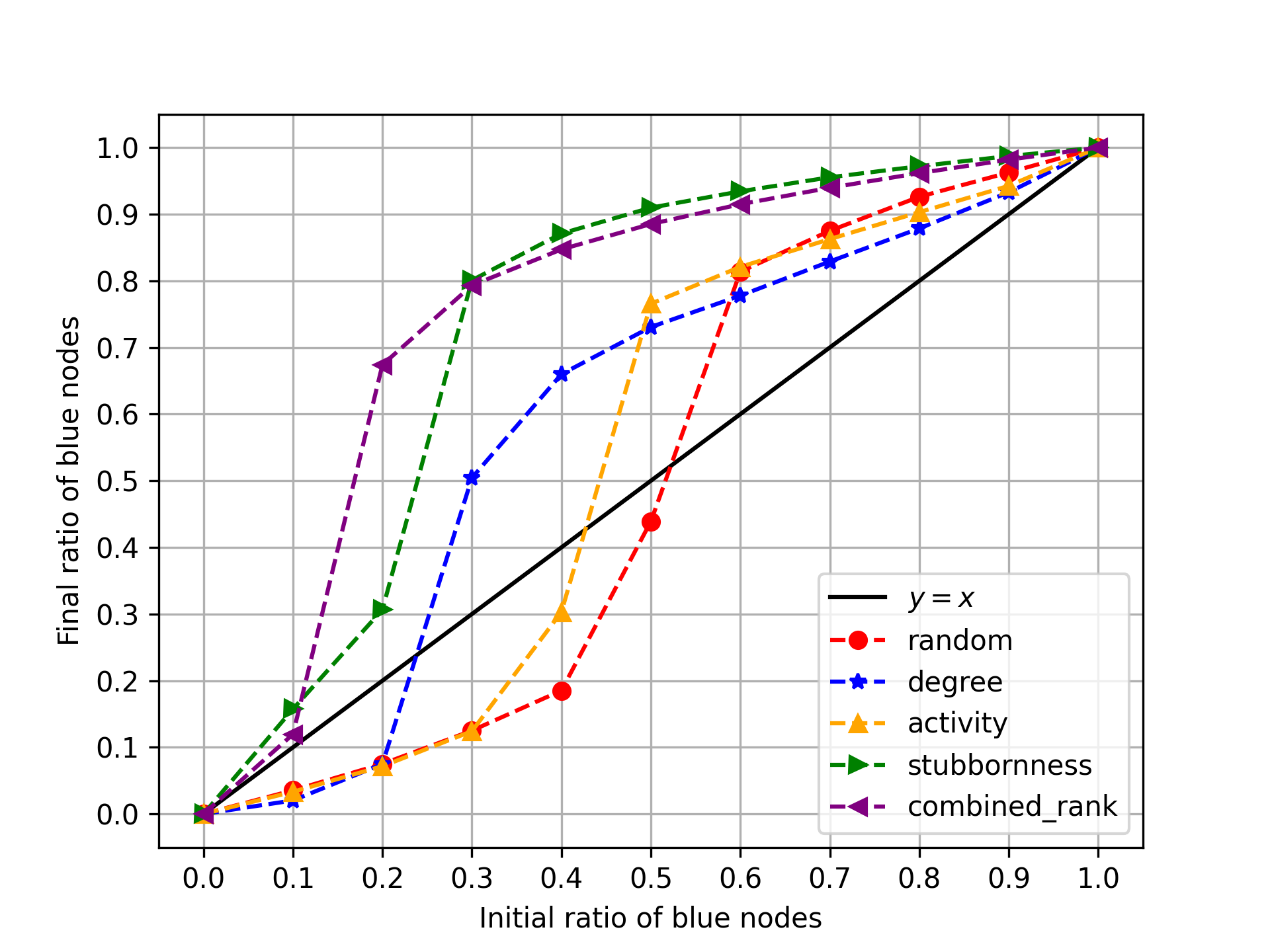}
         \captionsetup{justification=centering}
         \caption{}
         \label{fig9.1}
     \end{subfigure}
     \hfill
     \begin{subfigure}[b]{0.49\textwidth}
         \centering
         \includegraphics[width=\textwidth]{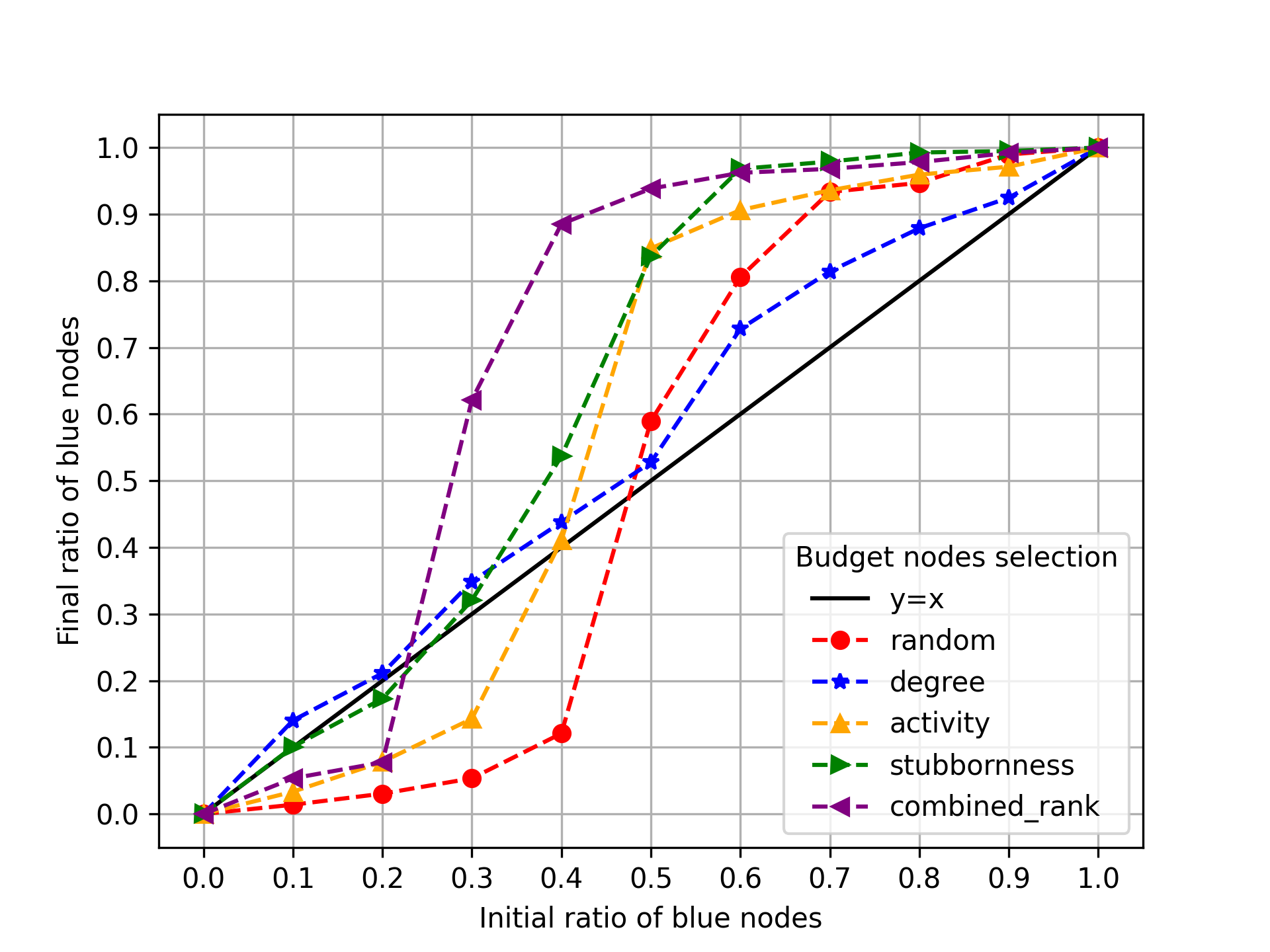}
         \captionsetup{justification=centering}
         \caption{}
         \label{fig9.2}
     \end{subfigure}
     \hfill
    \captionsetup{justification=justified}
    \caption{Personality-based influential nodes selection on (\textbf{a}) Twitch Spain's and (\textbf{b}) Facebook's social network.}
    \label{fig9}
\end{figure}

The results are illustrated in Figure~\ref{fig9}. It can be observed that selecting the initial blue nodes according to the stubbornness of nodes always maximizes the final ratio of blue nodes in T-ES social networks. In 
FB social network, the stubbornness-based strategy maximizes the final ratio of blue nodes when the initial ratio is greater than or equal to 0.4. 

\par\medskip

The results indicate that the stubbornness is the dominant parameter in this model. But, what about selecting the initial nodes based on a rank that combines the centrality and personality of a node? A new parameter called combined rank is introduced, considering $d(v)$, $act(v)$, and $stub(v)$. As stubbornness has more of an impact on the final ratio of blue nodes than the other two factors, more weight are added to the $stub(b)$ in the combined rank. Thus, the combined rank $r(v)$ is defined as:
$$r(v) = \sqrt{d(v)}\cdot \sqrt{act(v)}\cdot (stub(v))^3.$$

We ran another experiment in which the initial blue nodes were picked up based on the combined rank. The outcomes are the purple line in Figure~\ref{fig9.1} and Figure~\ref{fig9.2}. It can be observed that only 20\% of the initial blue nodes will result in about 70\% of blue nodes at the end in the T-ES social network. For FB social network, 30\% of the initial blue nodes result in about 60\% of blue nodes. 

\section{Conclusions}

In the current paper, we generalized the classic Galam opinion formation model by using a graph structure to represent a social network. Conducting several experiments on some special types of graphs, we showed that the convergence time of the GGM on the cycle and complete graphs are in $\mathcal{O}(n^2)$ and $\mathcal{O}(\log n)$, respectively, and it is probable that the expansion of a graph can influence the convergence time of this model. The model on a graph with strong expansion properties (e.g., any ``small'' subset of vertices of the graph has a great amount of connections with the complement of the subset, or the second-largest eigenvalue of the adjacency matrix of the graph is small) is expected to converge faster. Finding an explicit relationship between the expansion of a graph and the convergence time of GGM could be an interesting problem to tackle in the future.

Furthermore, experiments on real-world social networks indicate that selecting the initial blue nodes based on their centrality properties and personalities can lead to more final blue nodes compared with choosing the initial blue nodes randomly. It should be noted that the number of blue nodes just oscillated around the initial number when we select the initial blue nodes according to the closeness centrality on FB's social network. What types of graph structure can lead to such an outcome? Why does the structure not affect the outcomes of degree-based and betweenness-based strategies? This is left for future studies. 

In addition, selecting the initial blue nodes after considering all factors (using combined rank) can obtain even better results. A further study of such hybrid strategies can be a potential avenue for future research.   


\textbf{Acknowledgments: }We express our deepest appreciation to Serge Galam for his invaluable advice and feedback on this paper. This paper would not have been possible without his guidance and support.

\end{document}